\documentclass[aps,prl,twocolumn,showpacs,groupedaddress,amsmath,amssymb]{revtex4}

\usepackage{pslatex}
\usepackage{amsmath}
\usepackage{graphicx}

\begin{document}
\bibliographystyle{apsrev}

\title{$1\!/\!f$ Noise in Electron Glasses}
\author{Kirill Shtengel}
\email[]{shtengel@microsoft.com}
\altaffiliation{Microsoft, Redmond, WA 98052}
\affiliation{Department of Physics and Astronomy, University of
  California Irvine, Irvine, CA 92697-4575}

\author{Clare C. Yu}
\email[]{cyu@uci.edu}
\affiliation{Department of Physics and Astronomy, University of
  California Irvine, Irvine,
  CA 92697-4575}

\date{\today}

\begin{abstract}
  We show that $1/f$ noise is produced in a 3D electron glass by charge
  fluctuations due to electrons hopping between isolated sites and a
  percolating network at low temperatures. 
  The low frequency noise spectrum goes as $\omega^{-\alpha}$ with $\alpha$
  slightly larger than 1.  This result together with the temperature
  dependence of $\alpha$ and the noise amplitude are in good agreement
  with the recent experiments. These results hold true both with
  a flat, noninteracting density of states and with a density of states
  that includes Coulomb interactions. In the latter case, the density of
  states has a Coulomb gap that
  fills in with increasing temperature. For a large Coulomb gap width,
  this density of states gives
  a dc conductivity with a hopping exponent of $\approx 0.75$ 
  which has been observed in recent experiments. 
  For a small Coulomb gap width, the hopping exponent $\approx 0.5$.
\end{abstract}


\maketitle

\section{Introduction}
Low frequency $1/f$ noise \cite{Dutta81,Weissman88,Kogan96} is found
in a wide variety of conducting systems such as metals,
semiconductors, tunnel junctions \cite{Rogers84}, and even
superconducting SQUIDs \cite{Koch83,Koelle99}. Yet the microscopic
mechanisms are still not well understood. One example is an electron
glass which is an insulator where electrons are localized by a strong
random potential. A special case of this is a Coulomb glass in which
the electrons interact with one another via a long range Coulomb
potential.  Doped semiconductors and strongly disordered metals
provide examples of electron glasses.  Experimental studies on doped
silicon inversion layers have shown that low frequency $1/f$ noise is
produced by hopping conduction \cite{Voss78}. Because the systems are
glassy, electron hopping can occur on very long time scales which can
produce low frequency noise.  In this paper we show that the resulting
noise spectrum goes as $f^{-\alpha}$ where $f$ is frequency and the
temperature dependent exponent $\alpha > 1$.

Shklovskii has suggested that $1/f$ noise is caused by fluctuations in
the number of electrons in an infinite percolating cluster
\cite{Shklovskii80}. These fluctuations are caused by the slow
exchange of electrons between the infinite conducting cluster and
small isolated donor clusters. Subsequently Kogan and Shklovskii
combined a more rigorous calculation with numerical simulations and
found a noise spectrum where $\alpha$ was considerably lower than 1
\cite{Kogan81}. Furthermore, below a minimum frequency of order 1--100
Hz, the noise spectral density saturated and became a constant
independent of frequency. Their calculations were valid only in the
high temperature regime where the impurity band was assumed to be
occupied uniformly and long-range Coulomb correlations were
essentially neglected.  Since then there have been attempts to include
the effects of correlations.

In particular, Kozub suggested a model \cite{Kozub96} in which
electron hops within isolated pairs of impurities produce fluctuations
in the potential seen by other hopping electrons that contribute to
the current. While leading to $1/f$--type noise within some frequency
range, this model also shows low frequency noise saturation due to the
exponentially small probability of finding an \emph{isolated} pair of
sites with a long tunneling time.  Moreover, the noise magnitude is
predicted to increase as the temperature $T\to 0$ in contradiction
with the recent experimental findings of Massey and Lee
\cite{Massey97}. This, in part, led Massey and Lee to the conclusion
that the single particle picture is inconsistent with the observed
noise behavior.  A different approach was proposed by Kogan
\cite{Kogan98} who considered intervalley transitions as the source of
the hopping conduction noise.  Unfortunately this approach does not
seem to be analytically tractable and is not easily generalizable.

In this paper we extend Kogan and Shklovskii's approach \cite{Kogan81}
by including the energy dependence of the hopping as well as the
effects of electron--electron interactions on the single particle
density of states $g(\varepsilon)$.  This is essentially a mean field
approximation: we assume that charge is carried by electron-like
quasiparticles whose interaction with the other charges is taken into
account via the single particle density of states.  Later we will
present some justification for why we believe this approach works 
for low frequency noise. For comparison we also consider the case
of noninteracting electrons with a flat density of states.

The paper is organized as follows. In section IIa, we describe our
calculation of the noise spectral density. In section IIb, we present
the density of states that includes the Coulomb gap and that models
the decrease in the gap with increasing temperature. We show that this
form of the density of states yields the usual value of the hopping 
exponent $\delta\approx$ 0.5 for small values of the Coulomb gap 
width $E_g$. However, for large values of $E_g$, $\delta\approx$ 0.75.
Both values have been seen experimentally 
\cite{Markovic00,Putten92,vanKeuls97,Adkins98,Gershenson00,Massey97}.
In section III, we present our results.

\section{Calculation}
\subsection{Noise Spectral Density}
We start with a model of the Coulomb glass in which electrons occupy
half of the impurity sites. Each site can have at most 1 electron due
to a large onsite repulsion.  The sites are randomly placed according
to a uniform spatial distribution, and each has a random onsite energy
$\phi_{i}$ chosen from a uniform distribution extending from $-W/2$ to
$W/2$. Thus, $g_\mathrm{o}$, the density of states without interactions,
is flat.  At $T=0$ such a system is a perfect insulator while at low
but finite temperatures it will be able to conduct via variable range
hopping \cite{Mott68,Ambegaokar71,efrosbook}.
In this picture the DC conductivity is dominated by particles hopping
along the percolating network, which is constructed as follows.  The
resistance $R_{ij}$ associated with a transition between sites $i$ and
$j$ grows exponentially with both their separation $r_{ij}$ and energy
difference $\varepsilon_{ij}$:
\begin{equation}
  R_{ij}=R_{ij}^\mathrm{o}\exp(x_{ij})
  \label{eq:resistanceij}
\end{equation}
where the prefactor $R_{ij}^\mathrm{o}=kT/(e^{2}\gamma_{ij}^\mathrm{\,o})$
with $\gamma_{ij}^\mathrm{\,o}$ being given by \cite{efrosbook}
\begin{equation}
  \gamma_{ij}^\mathrm{\,o}=\frac{D^{2}|\Delta^{j}_{i}|}{\pi \rho
    s^{5}\hbar^{4}} \left[\frac{2e^{2}}{3\kappa
      \xi}\right]^{2}\frac{r_{ij}^{2}}{\xi^2}
  \left[1+\left(\frac{\Delta^{j}_{i}\xi}{2\hbar
        s}\right)^{2}\right]^{-4}
  \label{eq:gamma0}
\end{equation}
where $D$ is the deformation potential, $s$ is the speed of sound,
$\rho$ is the mass density, $\xi$ is the localization length and
$\kappa$ is the dielectric constant. $\Delta_{i}^{j} =
\varepsilon_{j}-\varepsilon_{i}-e^{2}/\kappa r_{ij}$ is the change in
energy that results from hopping from $i$ to $j$ with
$\varepsilon_{i}=\phi_{i}+\sum_{j} \frac{e^{2}}{\kappa r_{ij}}n_{j}$
being a single site energy. In Eq.~(\ref{eq:resistanceij}), the
exponent is given by
\begin{equation}
  x_{ij}=\frac{2r_{ij}}{\xi}+\frac{\varepsilon_{ij}}{kT}
  \label{eq:xij}
\end{equation}
The exponent reflects the thermally activated hopping rate between $i$
and $j$ as well as the wavefunction overlap between the sites.
\begin{equation}
  \varepsilon_{ij}=\left\{
    \begin{array}{ll}
      |\varepsilon_j-\varepsilon_i|-\frac{e^2}{\kappa r_{ij}}, &
      \mbox{$(\varepsilon_i-\mu)(\varepsilon_j-\mu)<0$}\\
      \mathrm{max}
      \left[|\varepsilon_{i}-\mu|,|\varepsilon_{j}-\mu| \right],
      &
      \mbox{$(\varepsilon_i-\mu)(\varepsilon_j-\mu)>0$}\\
    \end{array}
  \right.
  \label{eq:Eij}
\end{equation}
(In what follows we choose the Fermi level $\mu = 0$.)

A noninteracting picture of DC conduction is described in terms of
electron hopping between sites in a cluster that spans the entire
sample. In order to determine which sites are in a cluster, we
introduce the ``acceptance'' parameter $x$ such that any two sites $i$
and $j$ are considered ``connected'' if $x_{ij}\leq x$ and
disconnected otherwise. For small values of $x$ only rare pairs of
sites are connected. As we increase $x$, more such pairs appear and
small clusters start coalescing into bigger ones until an infinite
cluster -- the critical percolating network -- is formed at some
$x_\mathrm{c}$. At this point we can neglect the contribution of the
remaining impurity sites to the DC conductivity since it is
exponentially small compared to that of the sites already in the
percolating network (although the former sites are important for
understanding both AC conductivity and noise). In the same spirit, the
resistance of the critical percolating network is dominated by a few
pairs with $x_{ij} = x_\mathrm{c}$ -- these are the pairs that bridge
the gaps between large finite clusters enabling the formation of the
infinite cluster. Hence, the resistance of the entire sample is well
approximated by $R_\mathrm{tot} \approx
\overline{R^\mathrm{o}}\exp(x_\mathrm{c})$ where
$\overline{R^\mathrm{o}}\equiv kT/(e^{2}\overline{\gamma^\mathrm{\,o}})$
with $\overline{\gamma^\mathrm{\,o}}$ being the average value of
$\gamma_{ij}^\mathrm{\,o}$ given by Eq.~(\ref{eq:gamma0}).

In the presence of Coulomb interactions, there is no exact mapping of
transport onto a percolation picture. We nevertheless 
assume that upon diagonalizing the interacting Hamiltonian one finds
that charge carrying excitations are of a local nature, and so they
can be treated within the percolation picture as noninteracting
quasiparticles. The Coulomb interactions renormalize the
single-particle density of states which acquires a soft gap. We will
discuss this in more detail in the section on the density of states.
However, we will mention here that this approach appears to work well for DC
conduction and leads to a temperature dependence of the conductivity
\cite{Efros75,efrosbook,Meir96} which is distinctly different from the
noninteracting case and which agrees with experiment (see for example ref.
\cite{Lee_combined}).  However, the question about the validity of
this approach is still far from being settled -- see
\cite{Perez-Garrido97} for a different point of view.

In our treatment we will focus on the noise caused by quasiparticle
hopping between isolated clusters and the percolating network,
producing fluctuations of charge in the latter
\cite{Shklovskii80,Kogan81}. Let $N_{\mathcal{P}}$ be the average
number of such particles in the critical percolating network and
$\delta N_{\mathcal{P}}(t)$ be its time-dependent fluctuation.
Assuming that only stationary processes are involved (i.e.
$\langle\delta N_{\mathcal{P}}(t_2)\delta
N_{\mathcal{P}}(t_1)\rangle=f(t_2-t_1)$), we can use the
Wiener--Khintchine theorem \cite{Kogan96} to relate the noise spectral
density $S_I(\omega)$ of current fluctuations to the Fourier transform
of the autocorrelation function:
\begin{equation}
  \frac{S_{I}(\omega)}{I^2}=
  \frac{2 \langle\delta N_{\mathcal{P}}(t_2)
    \delta N_{\mathcal{P}}(t_1)\rangle_\omega}
  {N_{\mathcal{P}}^2}.
  \label{eq:noise_spectrum}
\end{equation}
where $I$ is the average current. The charge fluctuation
autocorrelation function can be expressed as a superposition of modes
$\alpha$, each of which relax exponentially with a characteristic time
$\tau_{\alpha}$. Thus the Fourier transform $\langle \ldots
\rangle_\omega$ of the autocorrelation function is a weighted sum over
Lorentzians \cite{Kogan81}.
\begin{equation}
  {\langle\delta N_{\mathcal{P}}(t_2)
    \delta N_{\mathcal{P}}(t_1)\rangle_\omega}
  = \frac {2 k T}{e^2} \sum_{\alpha \neq 0}
  \frac{\tau_\alpha}{1+\omega^2 \tau_\alpha^2}
  \left|\sum_{i \in \mathcal{P}} C_i \psi_\alpha(i)\right|^2
  \label{eq:autocorrelation}
\end{equation}
Here $C_i \equiv \left(e^2/kT \right) \; f_i(1-f_i)$ is the
``capacitance'' of site $i$ (with
$f_i=\left[\exp(\varepsilon_i/kT)+1\right]^{-1}$ being its equilibrium
occupancy) while $\tau_{\alpha}^{-1}$ and $\psi_\alpha(i)$ are the
$\alpha$-th eigenvalue and eigenvector of the following system of
linear equations:
\begin{equation}
  \sum_j R_{ij}^{-1}\left[\psi_\alpha(i)-\psi_\alpha(j)\right]
  = \tau_\alpha^{-1} C_i \psi_\alpha(j)
  \label{eq:linear_system}
\end{equation}
with $R_{ij}$ being the inter-site resistances given by
Eq.~(\ref{eq:resistanceij}). Since $R_{ij}^{-1}$ is proportional to
the hopping rate
$\tau_{ij}^{-1}=\gamma_{ij}^\mathrm{\,o}\exp\left(-x_{ij}\right)$ from
site $i$ to site $j$, eq. (\ref{eq:linear_system}) relates
$\tau_{ij}^{-1}$ to the relaxation rates $\tau_{\alpha}^{-1}$ of the
entire percolating network.  The sum over sites $i$ in
Eq.~(\ref{eq:autocorrelation}) runs only over those sites that belong
to the critical percolating network (CN) since only their occupancies
affect the current through the sample. The physical meaning of the
quantity $C_i \psi_\alpha(i)$ is that it is proportional to the
fluctuation $\delta f_i$ of the occupation of site $i$ and decays
exponentially with the associated time constant $\tau_\alpha$. The
eigenvectors satisfy the following conditions:
\begin{eqnarray}
&\sum_{i}& C_i \psi_\alpha(i)\psi_\beta^{*}(i) = \delta_{\alpha \beta} 
\label{eq:orthonormal}\\ 
&\sum_{\alpha}& C_i \psi_\alpha(i)\psi_\alpha^{*}(j) = \delta_{ij}\\  
&\sum_{i}& C_i \psi_\alpha(i) = 0 \quad\quad \forall \alpha \neq 0
\label{eq:chargecons}
\end{eqnarray}
The first condition states that the eigenfunctions are orthonormal;
the second states that the functions form a complete set. One of the
eigenfunctions is a constant which we take to be the one corresponding
to $\alpha=0$. This has the eigenvalue $\tau_0^{-1}=0$. 
Eq. (\ref{eq:chargecons}) is the orthonormalization condition between
this eigenstate and the others. It represents the fact that the
fluctuations in occupation represented by the $\alpha\neq 0$ modes
do not affect the total number of electrons on the impuritiy sites.
Thus the last equation is just the statement of overall charge conservation.  
We remark here
that Eqs.~(\ref{eq:linear_system}) are \emph{linear} only within the
assumption made earlier of noninteracting quasiparticles.  Otherwise the
$R_{ij}$ are \emph{not} constant coefficients; they depend on the
onsite energies, which in turn depend on the occupancies of other
sites.

Since we are interested in the modes that affect the charge in
the conducting network, we can replace the sum over $\alpha$ by a sum
over all finite clusters that coalesce with the infinite cluster as the 
acceptance parameter increases above $x_c$. In particular we can replace
the sum over $\alpha$ by an integral over $x$ and a sum over 
all finite clusters merging 
with the infinite cluster at a given value of $x$. With this in mind, 
we can evaluate Eq.~(\ref{eq:autocorrelation}) using 
Eqs.~(\ref{eq:orthonormal}) and (\ref{eq:chargecons}). 
For a single mode $\alpha$ the sum over
sites $i$ can be split into a sum over finite clusters (FC) and a sum over the
infinite cluster (IC). So we can write the normalization condition 
Eq.~(\ref{eq:orthonormal}) and the charge conservation condition
Eq.~(\ref{eq:chargecons}) as
\begin{eqnarray}
\sum_{m\in{\rm FC}}C_m\psi_{\alpha}^{2}(m) + 
\sum_{n\in{\rm IC}}C_n\psi_{\alpha}^2(n)=1 
\label{eq:norm}\\
\sum_{m\in{\rm FC}}C_m\psi_{\alpha}(m) +
\sum_{n\in{\rm IC}}C_n\psi_{\alpha}(n)=0 
\label{eq:cons}
\end{eqnarray}
Since the fast modes equilibrate the occupations of sites within
each cluster, the eigenfunctions do not depend on their site
indices within each cluster, i.e., 
$\psi_{\alpha}(m)=\psi_{\alpha,{\rm FC}}$, $\forall\; m\;\in$ FC and
$\psi_{\alpha}(n)=\psi_{\alpha,{\rm IC}}$, $\forall\; m\;\in$ IC.
As a result the we can take $\psi_{\alpha,{\rm FC}}$
and $\psi_{\alpha,{\rm IC}}$ out of the sums in 
Eqs. (\ref{eq:norm}) and (\ref{eq:cons}). The sum over capacitances
in the finite clusters will be much smaller than the sum over the
infinite cluster which implies that $\left(\psi_{\alpha,{\rm IC}}\right)^2$ 
is negligible in Eq. (\ref{eq:norm}). This leads to 
\begin{equation}
\psi_{\alpha,{\rm FC}}=\left(\sum_{m\in {\rm FC}}C_m\right)^{-1/2}
\end{equation}
Plugging this into Eq. (\ref{eq:cons}) yields
\begin{equation}
\psi_{\alpha,{\rm IC}}\sum_{m\in {\rm IC}}C_m=-\left(\sum_{m\in {\rm FC}}C_m\right)^{1/2}
\end{equation}
We can use these results to evaluate the sum over sites in Eq.~(\ref{eq:autocorrelation})
by noting that all the sites in the critical network are also in the infinite
cluster by definition. Thus
\begin{eqnarray}
\sum_{i \in {\mathcal{P}}}C_{i}\psi_{\alpha}(i)&=&
\frac{N_{\mathcal{P}}}{N_{\rm IC}(x)} 
\sum_{i\in {\rm IC}}C_{i}\psi_{\alpha,{\rm IC}}
\nonumber \\
&=& -\frac{N_{\mathcal{P}}}{N_{\rm IC}(x)} 
\left(\sum_{i\in{\rm FC}}C_i\right)^{1/2}
\end{eqnarray}
where $N_{\rm IC}(x)$ is the number of sites in the infinite cluster at a given
value of $x$.

In evaluating Eq.~(\ref{eq:autocorrelation}), we make the following
approximation for $\tau_{\alpha}$. Since we are interested in the
modes $\alpha$ that affect the charge of the percolating network, we
only consider particle exchange between the isolated clusters and the
infinite cluster.  This involves hopping times that are longer than
those within the percolating network itself by definition. Due to the
exponentially wide distribution of hopping times $\tau_{ij}$ such
exchange is likely to be dominated by the single closest pair of sites
of which one belongs to the finite, and the other to the infinite
cluster. The relaxation times within each cluster are much faster, and
therefore the above mentioned pair serves as a ``bottleneck'' for
intercluster relaxation. A simple diagonalization of the system of
equations~(\ref{eq:linear_system}) for two clusters $\mathcal{A}_1$
and $\mathcal{A}_2$, with the ``bottleneck'' hopping resistance
$R=\mathrm{min}(R_{ij};\; i \in \mathcal{A}_1,\; j \in \mathcal{A}_2)$
between them (and with the assumption that all other intercluster
resistances are much higher and all intracluster resistances are much
lower than $R$) leads to the following expression for the intercluster
relaxation time:
\begin{equation}
  \tau = R\left(\left[\sum_{i \in \mathcal{A}_1} C_i\right]^{-1}+
    \left[\sum_{j\in \mathcal{A}_2} C_j\right]^{-1}\right)^{-1}.
  \label{eq:intercluster}
\end{equation}
Since we are interested only in the
situation when one of the clusters is infinite, this simplifies
Eq.~(\ref{eq:intercluster}): $\tau = R \,\sum_{i \in \mathcal{A}}
C_i$, where $\mathcal{A}$ is the finite cluster.

We can substitute this value of $\tau$ into
Eq.~(\ref{eq:autocorrelation}) by replacing the sum over all modes
$\alpha$ by a sum over all finite clusters that coalesce with the
infinite cluster as the acceptance parameter $x$ is increased above
$x_\mathrm{c}$. Each such finite cluster contributes one new term to the
sum over $\alpha$ in Eq.~(\ref{eq:autocorrelation}) with the
corresponding $\tau_{\alpha} = R(x) \,\sum_{i \in \mathcal{A}} C_i$
where $R(x) ={\overline{R^\mathrm{o}}}\,\mathrm{e}^{x}$. 
Then we can write the spectral density of the noise as follows:
\begin{equation}
  \frac{S_{I}(\omega)}{I^2}
  = \frac {16 k T}{e^2}
  \int_{\lambda x_\mathrm{c}}^\infty \!\mathrm{d}x \,
  {\sum_{\mathcal{A}}}^\prime
  \frac{N_{\mathrm{IC}}^{-2}(x) R(x) \,
    \left(\sum_{i \in \mathcal{A}} C_i \right)^2}
  {1+\omega^2\,R^2(x) \,
    \left(\sum_{i \in \mathcal{A}} C_i \right)^2}
  \label{eq:noise1}
\end{equation}
where $\sum_{\mathcal{A}}^\prime$ stands for the sum over all
finite clusters that coalesce with the infinite cluster as $x$
increases by $\mathrm{d}x$. The parameter $\lambda\geq 1$ and sets the
distance in $x$ space from the percolation threshold.

This equation is difficult to evaluate mathematically. Fortunately,
however, we can extract the low frequency asymptotic behavior of
Eq.~(\ref{eq:noise1}) where the above approximations are well
justified.  The lowest frequency contributions come from large values
of $x$ where the infinite cluster has already absorbed almost all the
sites (i.e. $N_{\mathrm{IC}}\approx N$, the total number of sites).
What is left are the small clusters, which are mostly isolated
sites in the increasingly rare voids of the infinite cluster. The
probability of having two such sites in the same void is negligibly
small. Since low frequency noise will be dominated by the hops between
such isolated sites and the infinite cluster, we only consider such
hops in obtaining the spectral density of current fluctuations.
In Eq. (\ref{eq:noise1}) we can set $\lambda$ to correspond to this
situation at large $x$, and we can replace the sum over all finite
clusters that are merging with the infinite cluster with a sum 
over all sites multiplied by the probability 
$\tilde{P}_{1}(x,\varepsilon)dx$ that a single site with energy $\varepsilon$ 
has its nearest neighbor between $x$ and $x + dx$.

We can write down an expression for $\tilde{P}_{1}(x,\varepsilon)dx$. 
We begin
by defining $P_1(x,\varepsilon)$ to be the probability that a given site
with the onsite energy $\varepsilon$ has no neighbors nearer than $x$.
Let $\exp[-\rho(x,\varepsilon)dx]$ be the probability that a site with
energy $\varepsilon$ has no neighbors between $x$ and $x + dx$. Then
\begin{equation}
P_1(x,\varepsilon)=
\exp\left(-\int_{0}^{x}\rho(x^{\prime},\varepsilon)dx^{\prime}\right)
\end{equation}
We can use this to express $\tilde{P}_{1}(x,\varepsilon)dx$ as the product
of $P_1(x,\varepsilon)$, the probability of no neighbors within $x$,
multiplied by the probability of having a neighbor between $x$ and
$x+dx$:
\begin{eqnarray}
\tilde{P}_{1}(x,\varepsilon)dx&=&P_{1}(x,\varepsilon)\left[1 -
e^{-\rho(x,\varepsilon)dx}\right]\\
&=&-\left[\frac{\partial}{\partial x}P_{1}(x,\varepsilon)\right]dx
\end{eqnarray}
Thus $(-\partial P_1/\partial x)$ is the probability density for a
site to have its nearest neighbor between $x$ and $x+dx$.
We can now write the spectral density of current fluctuations as
\begin{eqnarray}
  \frac{S_{I}(\omega)}{I^2} = \frac {16 k T V}{e^2 N^2}
  \int_{\lambda x_\mathrm{c}}^\infty \mathrm{d}x 
  \int_{-W/2}^{W/2}\!\!\!\!\!\!\! & &\!\!\!\! \mathrm{d}\varepsilon \; 
  g(\varepsilon,T)\,
  \left(-\frac{\partial P_1(x,\varepsilon)}{\partial x}\right)
  \nonumber \\
  & \times & \!\! \frac{R(x) \, C^2(\varepsilon)}
  {1+\omega^2\,R^2(x) \, C^2(\varepsilon)}
  \label{eq:noise2}
\end{eqnarray}
where $V$ is the volume, $W$ is the bandwidth, and 
$f(\varepsilon)$ is the Fermi occupation number. 
To obtain an expression for $P_1(x,\varepsilon)$, we note that the
average number $dN$ of impurity sites found in a phase volume element
$d\Omega=d^{d}r d\varepsilon^{\prime}$ within a distance $x$ of
a site with energy $\varepsilon$ is given by
\begin{equation}
dN=g(\varepsilon^{\prime})\theta\left(x - \frac{2r}{\xi}-\frac{|\varepsilon|+
        |\varepsilon^{\prime}|+|\varepsilon-\varepsilon^{\prime}|}{2kT}
    \right)d\varepsilon^{\prime} d^{d}r
\end{equation}
The probability that no sites are in $d\Omega$ is given by
\begin{equation}
\lim_{N\rightarrow\infty}\left[1-\frac{dN}{N}\right]^{N}=
e^{-dN}
\end{equation}
Thus the probability $P_1(x,\varepsilon)$ that a given site
with the onsite energy $\varepsilon$ has no neighbors nearer than $x$
is given by
\begin{eqnarray}
  P_1(x,\varepsilon) & = & \exp\left\{
    -  \int \mathrm{d}^d r \;
    \int_{-W/2}^{W/2} \mathrm{d}\varepsilon^{\prime} \,
    g(\varepsilon^{\prime},T)\right.
  \nonumber \\
  & { } & \times \left.
    \theta \left(x - \frac{2r}{\xi}-\frac{|\varepsilon|+
        |\varepsilon^{\prime}|+|\varepsilon-\varepsilon^{\prime}|}{2kT}
    \right)\right\}.
  \label{eq:prob1}
\end{eqnarray}
Notice the absence of the Coulomb energy in the argument of the 
$\theta$-function in Eq.~(\ref{eq:prob1}), in accordance with our 
quasiparticle picture. 
Our quasiparticle picture is likely to work best for hops between 
isolated sites 
and the infinite cluster. Although one such hop may result in a 
sequence of other hops, these will mostly happen within the infinite 
cluster on a much shorter time-scale, effectively renormalizing the 
properties of the ``slow'' particle. As was mentioned earlier, these 
renormalizations can be included in the single particle density of states 
$g(\varepsilon,T)$. 

To facilitate evaluating the integral in Eq.~(\ref{eq:noise2}) 
numerically for the case where we include a Coulomb gap in the
density of states, we define the dimensionless variables 
$\tilde{r}= r /\xi$, $\tilde{\varepsilon}=\varepsilon/E_g$,
$\tilde{\omega}=\omega/\;\overline{\gamma^o}$, $\tilde{T}=kT/E_g$,
$\tilde{\tau}=\overline{\gamma^o}R(x)C(\varepsilon)=
f(\varepsilon)(1-f(\varepsilon))e^x$, and
$\tilde{g}(\tilde{\varepsilon},\tilde{T})=g(\varepsilon,T)/g_o$. 
$g_\mathrm{o}$ is the noninteracting density of states and
$E_\mathrm{g} \approx e^3 \sqrt{\pi g_\mathrm{o}/3\kappa^3}$ is the
characteristic width of the Coulomb gap.
Evaluating the integral over $x$ in Eq.~(\ref{eq:noise2}) leads us to define
\begin{equation}
\tilde{x}=2\tilde{r}+\frac{|\tilde{\varepsilon}|+|\tilde{\varepsilon}^{\prime}|
+|\tilde{\varepsilon}-\tilde{\varepsilon}^{\prime}|}{2\tilde{T}} 
\label{eq:xtilde}
\end{equation}
Then we can rewrite Eq. (\ref{eq:noise2}) as
\begin{eqnarray}
\frac{S_{I}(\omega)}{I^2}&=&A\int^{\tilde{W}/2}_{-\tilde{W}/2}
d\tilde{\varepsilon}\;\tilde{g}(\tilde{\varepsilon},\tilde{T})
\int^{\tilde{W}/2}_{-\tilde{W}/2}
d\tilde{\varepsilon}^{\prime}\tilde{g}(\tilde{\varepsilon}^{\prime},\tilde{T})
\int_{0}^{\tilde{R}_V}\tilde{r}^{2} d\tilde{r} \nonumber \\
&\times & \theta(\tilde{x}-\lambda x_c) 
\frac{P_{1}(\tilde{x},\tilde{\varepsilon})\tilde{\tau}
(\tilde{x},\tilde{\varepsilon})f(\tilde{\varepsilon})
\left[1-f(\tilde{\varepsilon})\right]}
{1+\tilde{\omega}^2\tilde{\tau}^2(\tilde{x},\tilde{\varepsilon})}
\label{eq:noiseDimensionless}
\end{eqnarray}
where $A=64\pi g^2_o E^2_g V\xi^3/(N^2\overline{\gamma^o})$, 
$\tilde{R_V}=(3V/4\pi)^{1/3}/\xi$, $\tilde{W}=W/E_g$, $\eta=4\pi g_oE_g\xi^3$, and
\begin{eqnarray}
P_{1}(\tilde{x},\tilde{\varepsilon})&=&\exp\left[-\eta
\int_{0}^{\tilde{R}_V}\tilde{r}^{\prime\;2} d\tilde{r}^{\prime}
\int^{\tilde{W}/2}_{-\tilde{W}/2} d\tilde{\varepsilon}^{\prime\prime}
\tilde{g}(\tilde{\varepsilon}^{\prime\prime},\tilde{T}) 
\right.  \nonumber\\
&  \times & \left.\theta\left(\tilde{x}-2\tilde{r}^{\prime}-
\frac{|\tilde{\varepsilon}|+|\tilde{\varepsilon}^{\prime\prime}|
+|\tilde{\varepsilon}-\tilde{\varepsilon}^{\prime\prime}|}{2\tilde{T}}\right)
\right]
\label{eq:P1Dimensionless}
\end{eqnarray}

For comparison we also consider the case with no Coulomb gap by setting
$g(\varepsilon,T)=g_o$ in Eqs. (\ref{eq:noise2}) and (\ref{eq:prob1}).
Since there is no natural energy scale, we do not rescale the energies.
However, we can define $\tilde{r}$, $\tilde{\tau}$, and $\tilde{\omega}$
as before. As a result, the definition of $\tilde{x}$
in Eq.~(\ref{eq:xtilde}) becomes
$\tilde{x}=2\tilde{r}+\left(|\varepsilon|+\left|\varepsilon^{\prime}\right|
+\left|\varepsilon-\varepsilon^{\prime}\right|\right)/(2T)$.
In Eq. (\ref{eq:noiseDimensionless}), $A$ is replaced by
$A_o={64 \pi V g^{2}_\mathrm{\!o}}\xi^{3}/{N^2
\overline{\gamma^\mathrm{\,o}}}$ and $\tilde{W}$ is replaced by simply
$W$. In Eq.~(\ref{eq:P1Dimensionless}) $\eta$ is replaced by
$\eta_o=4\pi\xi^3 g_o$. 

\section{Density of States}
At zero temperature, long-range interactions produce a Coulomb gap
centered at the Fermi energy in $g(\varepsilon,T)$
\cite{Pollak70,Efros75,Efros76,efrosbook}.  This gap arises because
the stability of the ground state with respect to single electron
hopping from an occupied site $i$ to an unoccupied site $j$ requires
that the energy difference $\Delta_{i}^{j}>0$.  At finite temperatures
the Coulomb gap is partially filled and the density of states no
longer vanishes at the Fermi energy
\cite{Levin87,Grannan93,Li94,Mogilyanskii89,Vojta93,Sarvestani95}.  
The exact form of $g(\varepsilon,T)$ is not known, but some have argued 
\cite{Mogilyanskii89,Vojta93,Sarvestani95} that its low temperature 
asymptotic behavior
is described by $g(\varepsilon=0,T) \sim T^{d-1}$.  We have done
Monte Carlo simulations of a three dimensional Coulomb glass with
off--diagonal disorder and we find that $g(\varepsilon=0,T)$ cannot
be described by a simple power law \cite{leewong,Grannan93}. The results of 
such simulations do not produce a density of states that is suitable 
for use in our noise integrals due to 
finite size effects. In particular $g(\varepsilon,T)$ goes to zero
at energies far away from the Fermi energy because of the finite 
size of the system.

\begin{figure}
    \includegraphics[width=3in]{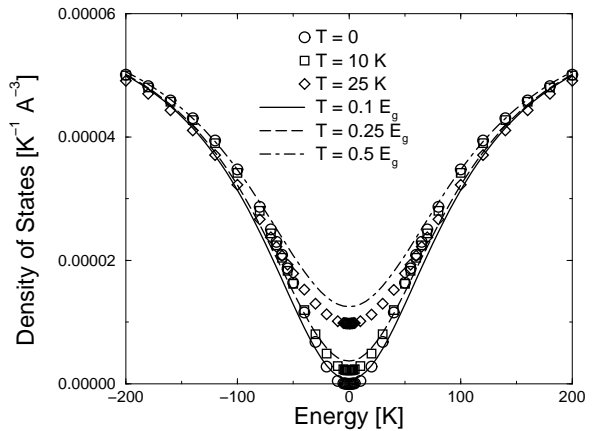}
    \caption{The density of states $g(\varepsilon,T)$ versus $\varepsilon$
      at various temperatures. The symbols are calculated using
      Eq.~(\ref{eq:BPW}) with $W_o/2=2.3\times 10^{4}$ K. 
      The density of states
      is measured from the Fermi energy $E_F=0$. The lines
      are the result of evaluating Eq.~(\ref{eq:dossimpleT}) with
      $E_g=100$ K.
      $g_o=6.25\times 10^{-5}$ states/K\AA$^3$. 
    }
    \label{fig:dos}
\end{figure}

Another way to approximate the density of states is to use the
Bethe--Peierls--Weiss (BPW) approximation \cite{Vojta93}.
The idea is to treat the interactions between one ``central'' site
and all other sites (boundary sites) exactly, but to include the
interactions between these boundary sites by means of effective
fields. The density of states can then be written as a convolution
\begin{equation}
g(\varepsilon,T)=\int_{-W_o/2}^{W_o/2}d\varepsilon^{\prime}
g\left(\varepsilon-\varepsilon^{\prime}\right)\frac{1}{kT}
\mathit{h}\left(\frac{\varepsilon^{\prime}}{kT}\right)
\label{eq:BPW}
\end{equation}
where $g(\varepsilon)$ is the zero temperature density of states
and $W_o$ is the bandwidth.
The function $\mathit{h}\left(\varepsilon/kT\right)$ takes into
account thermal fluctuations in the occupation of the
central site and the boundary sites. At low temperatures it has a sharp
peak with a width of the order $kT$ at $\varepsilon=0$.
We can make the approximation 
$(1/kT)\mathit{h}\left(\varepsilon/kT\right)\approx -f^{\prime}(\varepsilon)$
where $f^{\prime}(\varepsilon)$ is the derivative of the Fermi
function. The zero temperature density of states can be determined
numerically by solving a self--consistent equation based on
the ground state stability condition that a single electron
hopping from an occupied site $i$ to an unoccupied site $j$ requires
$\Delta^{j}_{i} > 0$ \cite{Baranovskii80,Yu99}. 
The result of evaluating Eq. (\ref{eq:BPW}) is shown in Fig.~\ref{fig:dos}.

Since using the BPW approximation to evaluate Eqs.~(\ref{eq:noise2})
and (\ref{eq:prob1}) is rather awkward, we model the finite
temperature density of states by
\begin{equation}
  g(\varepsilon,T)  =  g_\mathrm{o}\,\frac{ \varepsilon^2+(kT)^2}
    {E_\mathrm{g}^2 + \varepsilon^2 + (kT)^2}\,.
  \label{eq:dossimpleT}  
\end{equation}
Notice that for $T=0$,
$g(\varepsilon,T=0)\sim \varepsilon^{2}$ for $\varepsilon\ll E_g$ as 
is expected for a Coulomb gap in three dimensions. For large energies
($\varepsilon\gg E_g$ and $\varepsilon\gg kT$), 
$g(\varepsilon,T)$ approaches the noninteracting value $g_o$. A comparison
of Eq.~(\ref{eq:dossimpleT}) with the BPW approximation at various 
temperatures is shown in Fig.~\ref{fig:dos}. Eq.~(\ref{eq:dossimpleT})
is the expression we use for the density of states of a Coulomb glass
in Eqs.~(\ref{eq:noise2}) and (\ref{eq:prob1}).

\begin{figure}
    \includegraphics[width=3in]{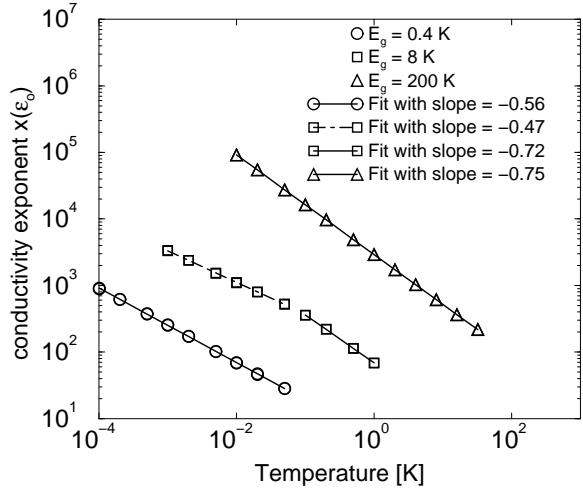}
    \caption{$x(\varepsilon_o,T)$ versus temperature with
     $g_o=6.25\times 10^{-5}$ states/K--$\AA^3$ and $\xi=10a_o=5.29177\;\AA$
     where $a_o$ is the Bohr radius and 10 is an estimate of
     the dielectric constant. We show plots for
     $E_g=0.4$ K ($\bigcirc$), $E_g=8$ K ($\square$) and 
     $E_g=200$ K ($\bigtriangleup$). $E_g=0.4$ K corresponds
     to the value of the Coulomb gap deduced from transport measurements
     while $E_g=8$ K value from tunneling measurements on Si:B 
     \cite{Meir96,Massey95}. The 
     lines are fits to the numerical data with the indicated slopes.
     The fit to the $E_g=0.4$ K data yields $\delta=0.56$ and $T_o=19$ K.
     The fit to the $E_g=8$ K data at low temperatures yields 
     $\delta=0.47$ and $T_o=27,206$ K, while the fit to the high
     temperature data yields  $\delta=0.72$ and $\delta=0.357$ K.
     The fit to the $E_g=200$ K data yields  $\delta=0.75$ and $T_o=42,068$ K.
     $\delta$ is virtually independent of $g_o$ but $T_o$ does depend on $g_o$. 
     For example, changing $g_o$ by 10 orders of magnitude to 
     $6.25\times 10^{+5}$ states/K--$\AA^3$ results in $\delta=0.75$
     and $T_o=19$ K for $E_g=200$ K.}
    \label{fig:xo}
\end{figure}

We can calculate the DC conductivity resulting from this density of
states by following Mott's argument for variable range hopping
\cite{efrosbook}. We start
with the hopping resistance $R_{ij}$ given by Eq.~(\ref{eq:resistanceij}).
Mott pointed out that hopping conduction at low temperatures comes
from states near the Fermi energy. If we consider states within 
$\varepsilon_o$ of the Fermi energy ($E_F=0$), then the 
concentration of states in this band is
\begin{equation}
N\left(\varepsilon_o,T\right)=\int^{\varepsilon_o}_{-\varepsilon_o}
g(\varepsilon,T)d\varepsilon
\end{equation}
where $g(\varepsilon,T)$ is given by Eq.~(\ref{eq:dossimpleT}).
So the typical separation between sites is 
$R_o=\left[N(\varepsilon_o,T\right]^{-1/3}$. To estimate the resistance
corresponding to hopping between two typical states in the band, we 
replace $r_{ij}$ with $R_o$ and $\varepsilon_{ij}$ with
$\varepsilon_o$ in Eq.~(\ref{eq:xij}) to obtain $x(\varepsilon_o)$.
Minimizing $x(\varepsilon_o)$ numerically yields $\varepsilon_o$. 
A plot of $x(\varepsilon_o)$ versus temperature is shown in Figure
\ref{fig:xo}. The dc
conductivity is then given by $\sigma(T)=\sigma_o\exp[-x(\varepsilon_o)]$.
We find that at low temperatures ($T\ll E_g$)
\begin{equation}
\sigma(T)=\sigma_o\exp\left[-\left(\frac{T_o}{T}\right)^{\delta}\right]
\end{equation}
where  $\delta$ is the hopping exponent. The value of $\delta$ depends
on $E_g$. For large values of the Coulomb gap ($E_g\stackrel{>}{\sim}50$ K)
$\delta\approx 0.75$ while for small values of the Coulomb gap
($E_g\stackrel{<}{\sim}1$ K) $\delta\approx 0.5$. When we tried
intermediate values of $T=8$, 10, and 20 K, we found that 
$\ln[x(\varepsilon_o)]$ versus $\ln(T)$ had a break in slope with
$\delta\approx 0.5$ at low temperatures and with $\delta\approx 0.72-0.75$
at high temperatures. Examples are shown in Fig.~\ref{fig:xo}. 
$\delta=0.75$ is higher
than the Mott value of $\delta=0.25$ associated with a flat density
of states and the value of $\delta=0.5$ derived by Efros and Shklovskii
\cite{Efros75} for the zero temperature Coulomb gap. However,
experiments on materials such as ultrathin metal films 
find values for $\delta=0.75\pm 0.05$ 
\cite{Markovic00,Putten92,vanKeuls97,Adkins98,Gershenson00} in
agreement with our value of $\delta$ for large $E_g$. The mechanism behind
this exponent has been a puzzle \cite{Markovic00,Phillips01}. 
Here we see that a possible simple
explanation for the experimental observation of an anomalous
hopping exponent is that the Coulomb gap in the single particle
density of states is filling in with increasing temperature.
If one takes this into account in the variable range hopping
calculations, then the observed exponent of 0.75 can be obtained naturally.
However, we should caution that our calculation applies to three dimensions
while a two dimensional calculation may be more appropriate for
ultrathin films. In fact we find that the analogous two dimensional
calculation with a density of states 
$g(\varepsilon,T)=g_o\left(|\varepsilon|+kT\right)/\left(E_g+|\varepsilon|+
kT\right)$ yields $\delta\approx$ 0.5.

\section{Results}
\begin{figure}
    \includegraphics[width=3.2in]{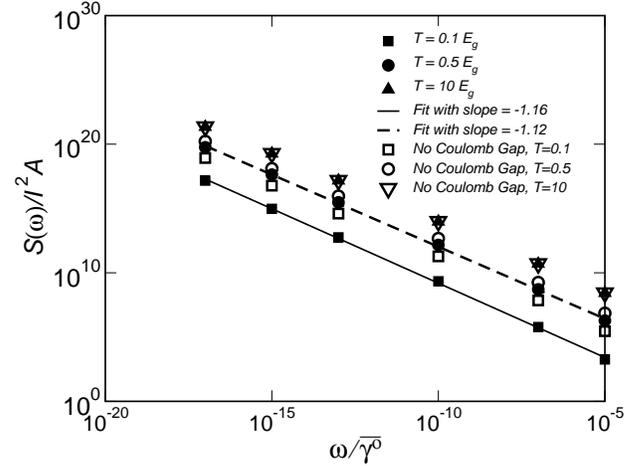}
    \caption{The noise power spectrum as a function of frequency. 
      The frequency is measured in the units of
      $\overline{\gamma^\mathrm{\,o}}$ which is estimated to be of the
      order $10^{13}$Hz for values appropriate for insulating Si:B.
      Unless otherwise noted,
      all curves in this and the following figures which were obtained for
      the case with a Coulomb gap used 
      $\eta=4\pi E_\mathrm{g} \xi^3 g_\mathrm{o} \sim 12
      [E_\mathrm{g}/(e^2/\kappa\xi)]^3 = 4.8\times 10^{-6}$, which in
      our estimates corresponds to the experimental dopant
      concentration of roughly $n=0.8\, n_\mathrm{c}$ for
      Si:B~\cite{Lee_combined,Massey97}. 
      We set $\tilde{W}=20$, $\tilde{R}_V=100$ and 
      $\lambda x_\mathrm{c}=1$ (the precise
      value of $\lambda$ has no effect on the low frequency noise
      which is governed by $x \gg x_\mathrm{c}$).  The parameter
      $A\equiv{64 \pi V E^{2}_\mathrm{g} g^{2}_\mathrm{\!o}}\xi^{3}/{N^2 
      \overline{\gamma^\mathrm{\,o}}}$. For comparison we show the 
      noise spectrum in
      the absence of a Coulomb gap with $g(\varepsilon,T)=g_o$ in
      Eqs. (\ref{eq:noise2}) and (\ref{eq:prob1}). In the absence of a 
      Coulomb gap, $A$ is replaced by $A_o\equiv
      {64 \pi V g^{2}_\mathrm{\!o}}\xi^{3}/{N^2
      \overline{\gamma^\mathrm{\,o}}}$ and $\eta$ is replaced by 
      $\eta_o=4\pi\xi^3 g_o=4.8\times10^{-6}$. The energy is measured
      in arbitrary units and we set $W=20$. The other variables are
      the same as in the case of a finite Coulomb gap.} 
    \label{fig:noise}
\end{figure}

\begin{figure}
    \includegraphics[width=3in]{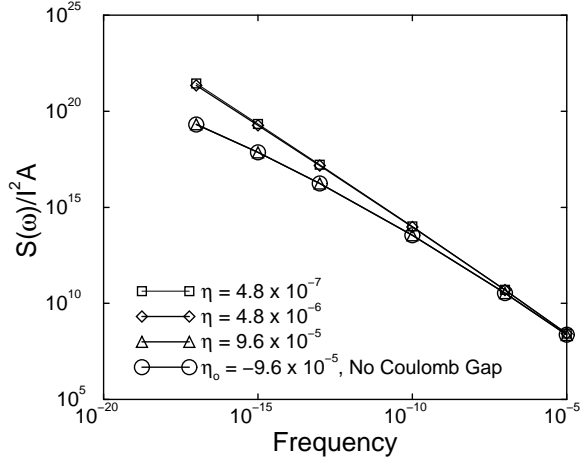}
    \caption{The noise power spectrum as a function of frequency at
      $T=10$ $E_g$ for various values of 
      $\eta=4\pi E_\mathrm{g} \xi^3 g_\mathrm{o}$. The rest of the
      parameters are the same as in Fig. \ref{fig:noise}. Notice
      the saturation at low frequencies for large $\eta$. For comparison
      we show the case with no Coulomb gap at $T=10$ with a large value of
      $\eta_o=4\pi\xi^3 g_o$. Large values of $\eta_o$ lead to saturation but
      small values do not.
}
\label{fig:EgVaries}
\end{figure}

\begin{figure}
    \includegraphics[width=3in]{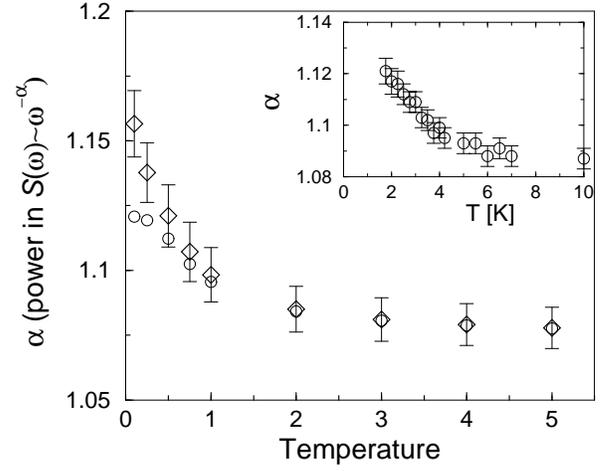}
    \caption{The spectral exponent $\alpha$ as a function
      of temperature with a Coulomb gap in the density of states
      ($\diamond$) and with a flat density of states ($\bigcirc$).
      We have suppressed the error bars for the case with no Coulomb
      gap to avoid cluttering the graph. The suppressed error bars
are comparable to those for the exponent with a Coulomb gap at high
temperatures. The temperature is measured in units of the Coulomb gap
$E_g$ for the case where there is a Coulomb gap, and in arbitrary
units for the case without a Coulomb gap.
      The inset shows the experimental data obtained
      for Si:B~\cite{Massey97}. }
    \label{fig:slope}
\end{figure}

\begin{figure}
    \includegraphics[width=3in]{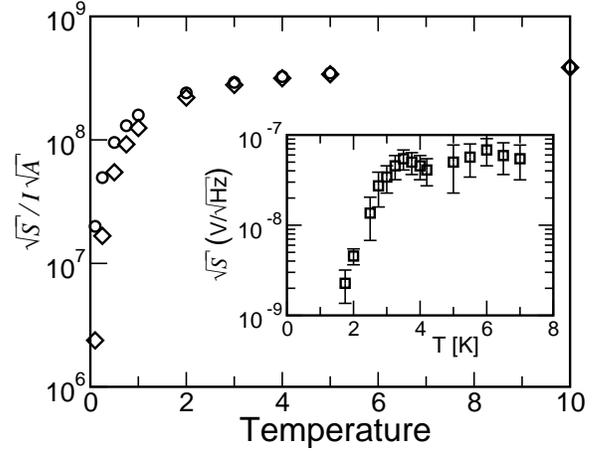}
    \caption{Noise amplitude $\sqrt{S}$ at
      $\omega = 10^{-13}\overline{\gamma^\mathrm{\,o}}$ (or $f \sim
      1$Hz) as a function of temperature for the cases
      with a Coulomb gap ($\diamond$) and without a Coulomb gap ($\bigcirc$).
      The temperature is measured in units of $E_g$ for the case of a
      finite Coulomb gap and in arbitrary units in the case of no Coulomb
      gap. The inset shows the
      experimental data for $f=1$Hz
      ~\cite{Massey97}.}
    \label{fig:amplitude}
\end{figure}

We evaluate Eqs.~(\ref{eq:noiseDimensionless}) and 
(\ref{eq:P1Dimensionless}) numerically and display
the results in Figs.~\ref{fig:noise}--\ref{fig:amplitude}. 
In Fig.~\ref{fig:noise} we show the spectral density of the noise
as a function of frequency. We find that for a
wide range of parameters the noise spectral density is given by
$S(\omega) \sim {\omega^{-\alpha}}$ with the spectral exponent
$\alpha$ between 1.07 and 1.16 (see Figs.~\ref{fig:noise},
\ref{fig:slope}) which is ``1/f'' noise.
For comparison we show in Fig. \ref{fig:noise} the noise spectrum in
the absence of a Coulomb gap with $g(\varepsilon,T)=g_o$ in
Eqs. (\ref{eq:noise2}) and (\ref{eq:prob1}).
The slope of a line through the open squares is $-1.12$ which is very close to
the values obtained with a Coulomb gap. Notice that the presence of
a Coulomb gap reduces the noise amplitude at low temperatures.

In Fig.~\ref{fig:noise} 
we use the \emph{transport} value of $E_\mathrm{g}\approx 0.4 K$, not the
tunneling one $\sim 8 K$; the two were found to be different by
an order of magnitude~\cite{Meir96,Lee_combined}. We
find that increasing $E_g$ by a factor of 20 
does not produce a noticable change of the results at low temperatures
($T=0.1$ $E_g$), but at high temperatures ($T=10$ $E_g$) 
it does lead to saturation of the noise power at low frequencies. 
This is shown in Figure \ref{fig:EgVaries} which also shows that
saturation occurs in the absence of a Coulomb gap when $\eta_o$ is
increased by a factor of 20. 
This saturation of the noise power occurs because the probability
$P_1(x,\varepsilon)$ of finding a site with no neighbors closer than
$x$ (see Eq.~(\ref{eq:prob1})) decreases exponentially with increasing
temperature and with increasing
$\eta$ or $\eta_o$. In addition $P_1(x,\varepsilon)$
becomes exponentially small as $x$
becomes large, and it is the large values of $x$ that contribute to
the low frequency noise. 
Finally we note that decreasing $E_g$ by a factor of 10 does not produce a 
noticable change of the results 
for either low temperatures ($T=0.1$ $E_g$) or high
temperatures ($T=10$ $E_g$).
We plot the spectral exponent $\alpha$ in Fig.~\ref{fig:slope}
versus temperature for the cases with and without a Coulomb
gap in the density of states. In both cases we see that it decreases slightly 
with increasing temperature and eventually saturates in 
qualitatively agreement with experiment \cite{Massey97}.
Fig.~\ref{fig:amplitude} shows that the noise amplitude $\sqrt{S}$
grows with temperature and eventually saturates, both in good
qualitative agreement with the experimental results of Massey and Lee
\cite{Massey97}. The data of Massey and Lee span 2 decades in
frequency while our calculations are able to cover a much broader range.
Again we see from Fig.~\ref{fig:amplitude} that
the presence of a Coulomb gap reduces the noise amplitude at low
temperatures. We obtain qualitatively the same results 
both with and without a Coulomb gap in the density of states which implies
that the behavior of the noise spectral density with respect to 
temperature and frequency is not strongly tied to the hopping exponent
$\delta$ or to the particular form of the density of states. 

We will now discuss some of the physical reasons behind our results. The fact
that we obtain $1/f$ noise is perhaps to be expected since weighted sums
over Lorentizians (see Eq.~(\ref{eq:autocorrelation})) often result
in $1/f$ noise \cite{Dutta81}. The subtlety lies in the temperature
dependence of the noise amplitude. For simplicity
let us consider the case of a density of states with no Coulomb gap which
gives qualitatively the same results as the case with a Coulomb gap. 
The decrease in the noise amplitude $\sqrt{S}$ with decreasing temperature 
is due to the presence of activated hopping processes which decrease with
decreasing temperature. However,
this is not at all obvious from Eq.~(\ref{eq:noiseDimensionless}). The integral
for the noise power at low frequencies is dominated by large $\tilde{x}$ which
corresponds to long relaxation times $\tilde{\tau}\sim\exp(\tilde{x})$. In this
case the factor of $f(\varepsilon)[1-f(\varepsilon)]$ cancels between the
numerator and denominator leaving the temperature dependence of the integrand
dominated by $P_1(x,\varepsilon)\exp(-\tilde{x})$. 
$P_1(x,\varepsilon)$ increases while $\exp(-\tilde{x})$ decreases with decreasing
temperature. The fact that our calculations yield an decrease
in the noise amplitude with decreasing temperature implies that the activated
hopping processes associated with $\exp(-\tilde{x})$ dominate. We should mention
that experimentally the noise power does not always decrease with 
decreasing temperature. In some cases it increases with decreasing 
temperature \cite{McCammon02,Shklovskii03} but we do not know the
differences in the samples which can account for this difference in behavior.

To summarize, recent experiments on $1/f$ noise~\cite{Massey97} are
consistent with a quasiparticle percolation picture of transport in 
electron glasses, though this does not exclude multi-particle
correlations.

\begin{acknowledgments}
  We would like to thank M.~Lee, M.~Pollak and M.~Weissman for useful
  and stimulating discussions. We thank Allen Goldman for bringing
  ref. \cite{Markovic00} to our attention. 
  This work was supported in part by ONR
  grant N00014-00-1-0005 and by DOE grant DE-FG03-00ER45843 as well as
  by the University of California Campus-Laboratory Collaborations
  program.
\end{acknowledgments}


\begin{thebibliography}{39}
\expandafter\ifx\csname natexlab\endcsname\relax\def\natexlab#1{#1}\fi
\expandafter\ifx\csname bibnamefont\endcsname\relax
  \def\bibnamefont#1{#1}\fi
\expandafter\ifx\csname bibfnamefont\endcsname\relax
  \def\bibfnamefont#1{#1}\fi
\expandafter\ifx\csname citenamefont\endcsname\relax
  \def\citenamefont#1{#1}\fi
\expandafter\ifx\csname url\endcsname\relax
  \def\url#1{\texttt{#1}}\fi
\expandafter\ifx\csname urlprefix\endcsname\relax\def\urlprefix{URL }\fi
\providecommand{\bibinfo}[2]{#2}
\providecommand{\eprint}[2][]{\url{#2}}

\bibitem[{\citenamefont{Dutta and Horn}(1981)}]{Dutta81}
\bibinfo{author}{\bibfnamefont{P.}~\bibnamefont{Dutta}} \bibnamefont{and}
  \bibinfo{author}{\bibfnamefont{P.~M.} \bibnamefont{Horn}},
  \bibinfo{journal}{Rev. Mod. Phys.} \textbf{\bibinfo{volume}{53}},
  \bibinfo{pages}{497} (\bibinfo{year}{1981}).

\bibitem[{\citenamefont{Weissman}(1988)}]{Weissman88}
\bibinfo{author}{\bibfnamefont{M.~B.} \bibnamefont{Weissman}},
  \bibinfo{journal}{Rev. Mod. Phys.} \textbf{\bibinfo{volume}{60}},
  \bibinfo{pages}{537} (\bibinfo{year}{1988}).

\bibitem[{\citenamefont{Kogan}(1996)}]{Kogan96}
\bibinfo{author}{\bibfnamefont{S.}~\bibnamefont{Kogan}},
  \emph{\bibinfo{title}{Electronic Noise and Fluctuations in Solids}}
  (\bibinfo{publisher}{Cambridge University Press},
  \bibinfo{address}{Cambridge}, \bibinfo{year}{1996}).

\bibitem[{\citenamefont{Rogers and Buhrman}(1984)}]{Rogers84}
\bibinfo{author}{\bibfnamefont{C.~T.} \bibnamefont{Rogers}} \bibnamefont{and}
  \bibinfo{author}{\bibfnamefont{R.~A.} \bibnamefont{Buhrman}},
  \bibinfo{journal}{Phys. Rev. Lett.} \textbf{\bibinfo{volume}{53}},
  \bibinfo{pages}{1272} (\bibinfo{year}{1984}).

\bibitem[{\citenamefont{Koch}(1983)}]{Koch83}
\bibinfo{author}{\bibfnamefont{R.~H.} \bibnamefont{Koch}}, in
  \emph{\bibinfo{booktitle}{Noise in Physical Systems and 1/f Noise}}, edited
  by \bibinfo{editor}{\bibfnamefont{M.}~\bibnamefont{Savelli}},
  \bibinfo{editor}{\bibfnamefont{G.}~\bibnamefont{Lecoy}}, \bibnamefont{and}
  \bibinfo{editor}{\bibfnamefont{J.-P.} \bibnamefont{Nougier}}
  (\bibinfo{publisher}{Elsevier Science Pub.}, \bibinfo{address}{Amsterdam},
  \bibinfo{year}{1983}), p. \bibinfo{pages}{377}.

\bibitem[{\citenamefont{Koelle et~al.}(1999)\citenamefont{Koelle, Kleiner,
  Ludwig, Dantsker, and Clarke}}]{Koelle99}
\bibinfo{author}{\bibfnamefont{D.}~\bibnamefont{Koelle}},
  \bibinfo{author}{\bibfnamefont{R.}~\bibnamefont{Kleiner}},
  \bibinfo{author}{\bibfnamefont{F.}~\bibnamefont{Ludwig}},
  \bibinfo{author}{\bibfnamefont{E.}~\bibnamefont{Dantsker}}, \bibnamefont{and}
  \bibinfo{author}{\bibfnamefont{J.}~\bibnamefont{Clarke}},
  \bibinfo{journal}{Rev. Mod. Phys.} \textbf{\bibinfo{volume}{71}},
  \bibinfo{pages}{631} (\bibinfo{year}{1999}).

\bibitem[{\citenamefont{Voss}(1978)}]{Voss78}
\bibinfo{author}{\bibfnamefont{R.~F.} \bibnamefont{Voss}}, \bibinfo{journal}{J.
  Phys. C} \textbf{\bibinfo{volume}{11}}, \bibinfo{pages}{L923}
  (\bibinfo{year}{1978}).

\bibitem[{\citenamefont{Shklovski\u{i}}(1980)}]{Shklovskii80}
\bibinfo{author}{\bibfnamefont{B.~I.} \bibnamefont{Shklovski\u{i}}},
  \bibinfo{journal}{Sol. St. Comm.} \textbf{\bibinfo{volume}{33}},
  \bibinfo{pages}{273} (\bibinfo{year}{1980}).

\bibitem[{\citenamefont{Kogan and Shklovksi\u{i}}(1981)}]{Kogan81}
\bibinfo{author}{\bibfnamefont{S.~M.} \bibnamefont{Kogan}} \bibnamefont{and}
  \bibinfo{author}{\bibfnamefont{B.~I.} \bibnamefont{Shklovksi\u{i}}},
  \bibinfo{journal}{Sov. Phys. Semicond.} \textbf{\bibinfo{volume}{15}},
  \bibinfo{pages}{605} (\bibinfo{year}{1981}).

\bibitem[{\citenamefont{Kozub}(1996)}]{Kozub96}
\bibinfo{author}{\bibfnamefont{V.~I.} \bibnamefont{Kozub}},
  \bibinfo{journal}{Sol. St. Comm.} \textbf{\bibinfo{volume}{97}},
  \bibinfo{pages}{843} (\bibinfo{year}{1996}).

\bibitem[{\citenamefont{Massey and Lee}(1997)}]{Massey97}
\bibinfo{author}{\bibfnamefont{J.~G.} \bibnamefont{Massey}} \bibnamefont{and}
  \bibinfo{author}{\bibfnamefont{M.}~\bibnamefont{Lee}},
  \bibinfo{journal}{Phys. Rev. Lett.} \textbf{\bibinfo{volume}{79}},
  \bibinfo{pages}{3986} (\bibinfo{year}{1997}).

\bibitem[{\citenamefont{Kogan}(1998)}]{Kogan98}
\bibinfo{author}{\bibfnamefont{S.}~\bibnamefont{Kogan}},
  \bibinfo{journal}{Phys. Rev. B} \textbf{\bibinfo{volume}{57}},
  \bibinfo{pages}{9736} (\bibinfo{year}{1998}).

\bibitem[{\citenamefont{Markovi\'c et~al.}(2000)\citenamefont{Markovi\'c,
  Christiansen, Grupp, Mack, Martinez-Arizala, and Goldman}}]{Markovic00}
\bibinfo{author}{\bibfnamefont{N.}~\bibnamefont{Markovi\'c}},
  \bibinfo{author}{\bibfnamefont{C.}~\bibnamefont{Christiansen}},
  \bibinfo{author}{\bibfnamefont{D.~E.} \bibnamefont{Grupp}},
  \bibinfo{author}{\bibfnamefont{A.~M.} \bibnamefont{Mack}},
  \bibinfo{author}{\bibfnamefont{G.}~\bibnamefont{Martinez-Arizala}},
  \bibnamefont{and} \bibinfo{author}{\bibfnamefont{A.~M.}
  \bibnamefont{Goldman}}, \bibinfo{journal}{Phys. Rev. B}
  \textbf{\bibinfo{volume}{62}}, \bibinfo{pages}{2195} (\bibinfo{year}{2000}),
  \bibinfo{note}{and references therein.}

\bibitem[{\citenamefont{van~der Putten et~al.}(1992)\citenamefont{van~der
  Putten, Moonen, Brom, Brokken-Zijp, and Michels}}]{Putten92}
\bibinfo{author}{\bibfnamefont{D.}~\bibnamefont{van~der Putten}},
  \bibinfo{author}{\bibfnamefont{J.~T.} \bibnamefont{Moonen}},
  \bibinfo{author}{\bibfnamefont{H.~B.} \bibnamefont{Brom}},
  \bibinfo{author}{\bibfnamefont{J.~C.~M.} \bibnamefont{Brokken-Zijp}},
  \bibnamefont{and} \bibinfo{author}{\bibfnamefont{M.~A.~J.}
  \bibnamefont{Michels}}, \bibinfo{journal}{Phys. Rev. Lett.}
  \textbf{\bibinfo{volume}{69}}, \bibinfo{pages}{494} (\bibinfo{year}{1992}).

\bibitem[{\citenamefont{Keuls et~al.}(1997)\citenamefont{Keuls, Hu, Jiang, and
  Dahm}}]{vanKeuls97}
\bibinfo{author}{\bibfnamefont{F.~W.~V.} \bibnamefont{Keuls}},
  \bibinfo{author}{\bibfnamefont{X.~L.} \bibnamefont{Hu}},
  \bibinfo{author}{\bibfnamefont{H.~W.} \bibnamefont{Jiang}}, \bibnamefont{and}
  \bibinfo{author}{\bibfnamefont{A.~J.} \bibnamefont{Dahm}},
  \bibinfo{journal}{Phys. Rev. B} \textbf{\bibinfo{volume}{56}},
  \bibinfo{pages}{1161} (\bibinfo{year}{1997}).

\bibitem[{\citenamefont{Adkins and Astrakharchik}(1998)}]{Adkins98}
\bibinfo{author}{\bibfnamefont{C.~J.} \bibnamefont{Adkins}} \bibnamefont{and}
  \bibinfo{author}{\bibfnamefont{E.~G.} \bibnamefont{Astrakharchik}},
  \bibinfo{journal}{J. Phys: Condens. Matter} \textbf{\bibinfo{volume}{10}},
  \bibinfo{pages}{6651} (\bibinfo{year}{1998}).

\bibitem[{\citenamefont{Gershenson et~al.}(2000)\citenamefont{Gershenson,
  Khavin, Reuter, Schafmeister, and Wieck}}]{Gershenson00}
\bibinfo{author}{\bibfnamefont{M.~E.} \bibnamefont{Gershenson}},
  \bibinfo{author}{\bibfnamefont{Y.~B.} \bibnamefont{Khavin}},
  \bibinfo{author}{\bibfnamefont{D.}~\bibnamefont{Reuter}},
  \bibinfo{author}{\bibfnamefont{P.}~\bibnamefont{Schafmeister}},
  \bibnamefont{and} \bibinfo{author}{\bibfnamefont{A.~D.} \bibnamefont{Wieck}},
  \bibinfo{journal}{Phys. Rev. Lett.} \textbf{\bibinfo{volume}{85}},
  \bibinfo{pages}{1718} (\bibinfo{year}{2000}).

\bibitem[{\citenamefont{Shklovskii and Efros}(1984)}]{efrosbook}
\bibinfo{author}{\bibfnamefont{B.~I.} \bibnamefont{Shklovskii}}
  \bibnamefont{and} \bibinfo{author}{\bibfnamefont{A.~L.} \bibnamefont{Efros}},
  \emph{\bibinfo{title}{Electronic Properties of Doped Semiconductors}}
  (\bibinfo{publisher}{Spinger-Verlag}, \bibinfo{address}{Berlin},
  \bibinfo{year}{1984}), \bibinfo{note}{and references therein}.

\bibitem[{\citenamefont{Mott}(1968)}]{Mott68}
\bibinfo{author}{\bibfnamefont{N.~F.} \bibnamefont{Mott}}, \bibinfo{journal}{J.
  Non--Cryst. Solids} \textbf{\bibinfo{volume}{1}}, \bibinfo{pages}{1}
  (\bibinfo{year}{1968}).

\bibitem[{\citenamefont{Ambegaokar et~al.}(1971)\citenamefont{Ambegaokar,
  Halperin, and Langer}}]{Ambegaokar71}
\bibinfo{author}{\bibfnamefont{V.}~\bibnamefont{Ambegaokar}},
  \bibinfo{author}{\bibfnamefont{B.~I.} \bibnamefont{Halperin}},
  \bibnamefont{and} \bibinfo{author}{\bibfnamefont{J.~S.}
  \bibnamefont{Langer}}, \bibinfo{journal}{Phys. Rev. B}
  \textbf{\bibinfo{volume}{4}}, \bibinfo{pages}{2612} (\bibinfo{year}{1971}).

\bibitem[{\citenamefont{Efros and Shklovskii}(1975)}]{Efros75}
\bibinfo{author}{\bibfnamefont{A.~L.} \bibnamefont{Efros}} \bibnamefont{and}
  \bibinfo{author}{\bibfnamefont{B.~I.} \bibnamefont{Shklovskii}},
  \bibinfo{journal}{J. Phys. C} \textbf{\bibinfo{volume}{8}},
  \bibinfo{pages}{L49} (\bibinfo{year}{1975}).

\bibitem[{\citenamefont{Meir}(1996)}]{Meir96}
\bibinfo{author}{\bibfnamefont{Y.}~\bibnamefont{Meir}}, \bibinfo{journal}{Phys.
  Rev. Lett.} \textbf{\bibinfo{volume}{77}}, \bibinfo{pages}{5265}
  (\bibinfo{year}{1996}).

\bibitem[{\citenamefont{Massey and Lee}(1995{\natexlab{a}})}]{Lee_combined}
\bibinfo{author}{\bibfnamefont{J.~G.} \bibnamefont{Massey}} \bibnamefont{and}
  \bibinfo{author}{\bibfnamefont{M.}~\bibnamefont{Lee}},
  \bibinfo{journal}{Phys. Rev. Lett.} \textbf{\bibinfo{volume}{75}},
  \bibinfo{pages}{4266} (\bibinfo{year}{1995}{\natexlab{a}}),
  \bibinfo{note}{\emph{ibid} \textbf{76}, 3399 (1996); Phys.\ Rev.\ B
  \textbf{62}, R13 270 (2000); M.\ Lee and M.\ L.\ Stutzmann, Phys. Rev. Lett.
  \textbf{87}, 056402 (2001); also M.~Lee, private communication}.

\bibitem[{\citenamefont{Perez-Garrido et~al.}(1997)\citenamefont{Perez-Garrido,
  Ortuno, Cuevas, Ruiz, and Pollak}}]{Perez-Garrido97}
\bibinfo{author}{\bibfnamefont{A.}~\bibnamefont{Perez-Garrido}},
  \bibinfo{author}{\bibfnamefont{M.}~\bibnamefont{Ortuno}},
  \bibinfo{author}{\bibfnamefont{E.}~\bibnamefont{Cuevas}},
  \bibinfo{author}{\bibfnamefont{J.}~\bibnamefont{Ruiz}}, \bibnamefont{and}
  \bibinfo{author}{\bibfnamefont{M.}~\bibnamefont{Pollak}},
  \bibinfo{journal}{Phys. Rev. B} \textbf{\bibinfo{volume}{55}},
  \bibinfo{pages}{R8630} (\bibinfo{year}{1997}).

\bibitem[{\citenamefont{Pollak}(1970)}]{Pollak70}
\bibinfo{author}{\bibfnamefont{M.}~\bibnamefont{Pollak}},
  \bibinfo{journal}{Disc. Faraday Soc.} \textbf{\bibinfo{volume}{50}},
  \bibinfo{pages}{13} (\bibinfo{year}{1970}).

\bibitem[{\citenamefont{Efros}(1976)}]{Efros76}
\bibinfo{author}{\bibfnamefont{A.~L.} \bibnamefont{Efros}},
  \bibinfo{journal}{J. Phys. C} \textbf{\bibinfo{volume}{9}},
  \bibinfo{pages}{2021} (\bibinfo{year}{1976}).

\bibitem[{\citenamefont{Levin et~al.}(1987)\citenamefont{Levin, Nguyen,
  Shklovski\u{i}, and \'{E}fros}}]{Levin87}
\bibinfo{author}{\bibfnamefont{E.~I.} \bibnamefont{Levin}},
  \bibinfo{author}{\bibfnamefont{V.~L.} \bibnamefont{Nguyen}},
  \bibinfo{author}{\bibfnamefont{B.~I.} \bibnamefont{Shklovski\u{i}}},
  \bibnamefont{and} \bibinfo{author}{\bibfnamefont{A.~L.}
  \bibnamefont{\'{E}fros}}, \bibinfo{journal}{Sov. Phys. JETP}
  \textbf{\bibinfo{volume}{65}}, \bibinfo{pages}{842} (\bibinfo{year}{1987}).

\bibitem[{\citenamefont{Grannan and Yu}(1993)}]{Grannan93}
\bibinfo{author}{\bibfnamefont{E.~R.} \bibnamefont{Grannan}} \bibnamefont{and}
  \bibinfo{author}{\bibfnamefont{C.~C.} \bibnamefont{Yu}},
  \bibinfo{journal}{Phys. Rev. Lett.} \textbf{\bibinfo{volume}{71}},
  \bibinfo{pages}{3335} (\bibinfo{year}{1993}).

\bibitem[{\citenamefont{Li and Phillips}(1994)}]{Li94}
\bibinfo{author}{\bibfnamefont{Q.}~\bibnamefont{Li}} \bibnamefont{and}
  \bibinfo{author}{\bibfnamefont{P.}~\bibnamefont{Phillips}},
  \bibinfo{journal}{Phys. Rev. B} \textbf{\bibinfo{volume}{49}},
  \bibinfo{pages}{10269} (\bibinfo{year}{1994}).

\bibitem[{\citenamefont{Mogilyanski\u{i} and Ra\u{i}kh}(1989)}]{Mogilyanskii89}
\bibinfo{author}{\bibfnamefont{A.~A.} \bibnamefont{Mogilyanski\u{i}}}
  \bibnamefont{and} \bibinfo{author}{\bibfnamefont{M.~E.}
  \bibnamefont{Ra\u{i}kh}}, \bibinfo{journal}{Sov. Phys. JETP}
  \textbf{\bibinfo{volume}{68}}, \bibinfo{pages}{1081} (\bibinfo{year}{1989}).

\bibitem[{\citenamefont{Vojta et~al.}(1993)\citenamefont{Vojta, John, and
  Schreiber}}]{Vojta93}
\bibinfo{author}{\bibfnamefont{T.}~\bibnamefont{Vojta}},
  \bibinfo{author}{\bibfnamefont{W.}~\bibnamefont{John}}, \bibnamefont{and}
  \bibinfo{author}{\bibfnamefont{M.}~\bibnamefont{Schreiber}},
  \bibinfo{journal}{J. Phys.: Condens. Matter} \textbf{\bibinfo{volume}{5}},
  \bibinfo{pages}{4989} (\bibinfo{year}{1993}).

\bibitem[{\citenamefont{Sarvestani et~al.}(1995)\citenamefont{Sarvestani,
  Schreiber, and Vojta}}]{Sarvestani95}
\bibinfo{author}{\bibfnamefont{M.}~\bibnamefont{Sarvestani}},
  \bibinfo{author}{\bibfnamefont{M.}~\bibnamefont{Schreiber}},
  \bibnamefont{and} \bibinfo{author}{\bibfnamefont{T.}~\bibnamefont{Vojta}},
  \bibinfo{journal}{Phys. Rev. B} \textbf{\bibinfo{volume}{52}},
  \bibinfo{pages}{R3820} (\bibinfo{year}{1995}).

\bibitem[{lee()}]{leewong}
\bibinfo{note}{M. H. Overlin, L. Wong, C. C. Yu, unpublished.}

\bibitem[{\citenamefont{Baranovski\u{i}
  et~al.}(1980)\citenamefont{Baranovski\u{i}, Shklovski\u{i}, and
  \'{E}fros}}]{Baranovskii80}
\bibinfo{author}{\bibfnamefont{S.~D.} \bibnamefont{Baranovski\u{i}}},
  \bibinfo{author}{\bibfnamefont{B.~I.} \bibnamefont{Shklovski\u{i}}},
  \bibnamefont{and} \bibinfo{author}{\bibfnamefont{A.~L.}
  \bibnamefont{\'{E}fros}}, \bibinfo{journal}{Sov. Phys. JETP}
  \textbf{\bibinfo{volume}{51}}, \bibinfo{pages}{199} (\bibinfo{year}{1980}).

\bibitem[{\citenamefont{Yu}(1999)}]{Yu99}
\bibinfo{author}{\bibfnamefont{C.~C.} \bibnamefont{Yu}},
  \bibinfo{journal}{Phys. Rev. Lett.} \textbf{\bibinfo{volume}{82}},
  \bibinfo{pages}{4074} (\bibinfo{year}{1999}), \bibinfo{note}{to determine the
  equilibrium value of $g(\varepsilon,T=0)$, we use the infinite time limit.
  The parameters are the same as given in this reference but with
  $g_o=6.25\times 10^{-5}$ states/K$\AA^3$.}

\bibitem[{\citenamefont{Phillips}(2001)}]{Phillips01}
\bibinfo{author}{\bibfnamefont{J.~C.} \bibnamefont{Phillips}},
  \bibinfo{journal}{Phys. Rev. B} \textbf{\bibinfo{volume}{64}},
  \bibinfo{pages}{035411} (\bibinfo{year}{2001}).

\bibitem[{\citenamefont{Massey and Lee}(1995{\natexlab{b}})}]{Massey95}
\bibinfo{author}{\bibfnamefont{J.~G.} \bibnamefont{Massey}} \bibnamefont{and}
  \bibinfo{author}{\bibfnamefont{M.}~\bibnamefont{Lee}},
  \bibinfo{journal}{Phys. Rev. Lett.} \textbf{\bibinfo{volume}{75}},
  \bibinfo{pages}{4266} (\bibinfo{year}{1995}{\natexlab{b}}).

\bibitem[{\citenamefont{McCammon et~al.}(2002)\citenamefont{McCammon, Galeazzi,
  Liu, Sanders, Smith, Tan, Boyce, Brekosky, Gygax, Kelley
  et~al.}}]{McCammon02}
\bibinfo{author}{\bibfnamefont{D.}~\bibnamefont{McCammon}},
  \bibinfo{author}{\bibfnamefont{M.}~\bibnamefont{Galeazzi}},
  \bibinfo{author}{\bibfnamefont{D.}~\bibnamefont{Liu}},
  \bibinfo{author}{\bibfnamefont{W.~T.} \bibnamefont{Sanders}},
  \bibinfo{author}{\bibfnamefont{B.}~\bibnamefont{Smith}},
  \bibinfo{author}{\bibfnamefont{P.}~\bibnamefont{Tan}},
  \bibinfo{author}{\bibfnamefont{K.~R.} \bibnamefont{Boyce}},
  \bibinfo{author}{\bibfnamefont{R.}~\bibnamefont{Brekosky}},
  \bibinfo{author}{\bibfnamefont{J.~D.} \bibnamefont{Gygax}},
  \bibinfo{author}{\bibfnamefont{R.}~\bibnamefont{Kelley}},
  \bibnamefont{et~al.}, \bibinfo{journal}{Phys. Stat. Sol. (b)}
  \textbf{\bibinfo{volume}{230}}, \bibinfo{pages}{197} (\bibinfo{year}{2002}).

\bibitem[{\citenamefont{Shklovskii}(2003)}]{Shklovskii03}
\bibinfo{author}{\bibfnamefont{B.~I.} \bibnamefont{Shklovskii}},
  \bibinfo{journal}{Phys. Rev. B} \textbf{\bibinfo{volume}{67}},
  \bibinfo{pages}{045201} (\bibinfo{year}{2003}).


\end{thebibliography}
\end{document}